\documentclass[12pt,aps,prb]{revtex4}
\usepackage{epsfig}
\usepackage{amsmath}

\normalsize

\def\ba{{\bf a}}

\def\bi{{\bf i}}
\def\bj{{\bf j}}
\def\bk{{\bf k}}

\def\bQ{{\bf Q}}

\def\b0{{\bf 0}}

\def\cM{{\cal M}}

\def\cU{{\cal U}}
\def\cV{{\cal V}}

\def\tba{\tilde\ba}

\def\bra{\langle}
\def\ket{\rangle}
\def\up{\uparrow}
\def\down{\downarrow}

\def\alf{\alpha}
\def\eps{\epsilon}

\def\Gam{\Gamma}
\def\lam{\lambda}
\def\Lam{\Lambda}

\def\Om{\Omega}
\def\sg{\sigma}

\def\det{{\rm det}}

\begin{document}


\title{Renormalized mean-field analysis of antiferromagnetism \\
and d-wave superconductivity in the two-dimensional \\ 
Hubbard model}

\author{J.\ Reiss$^1$, D.\ Rohe$^2$, and W.\ Metzner$^1$}

\affiliation{$^1$Max-Planck-Institute for Solid State Research, 
 D-70569 Stuttgart, Germany \\
 $^2$Centre de Physique Th\'eorique, Ecole Polytechnique,
 CNRS-UMR 7644, 91128 Palaiseau Cedex, France}

\date{\small\today}


\begin{abstract}
We analyze the competition between antiferromagnetism and 
superconductivity in the two-dimensional Hubbard model by combining
a functional renormalization group flow with a mean-field theory
for spontaneous symmetry breaking.
Effective interactions are computed by integrating out states
above a scale $\Lam_{\rm MF}$ in one-loop approximation, which
captures in particular the generation of an attraction in the
d-wave Cooper channel from fluctuations in the particle-hole channel.
These effective interactions are then used as an input for a 
mean-field treatment of the remaining low-energy states, with
antiferromagnetism, singlet superconductivity and triplet 
$\pi$-pairing as the possible order parameters.
Antiferromagnetism and superconductivity suppress each other,
leaving only a small region in parameter space where both orders
can coexist with a sizable order parameter for each. Triplet
$\pi$-pairing appears generically in the coexistence region, but
its feedback on the other order parameters is very small.\\
\noindent
\mbox{PACS: 71.10.Fd, 74.20.-z, 75.10.-b}
\end{abstract}


\maketitle


\section{Introduction}

Soon after the discovery of high-temperature superconductivity
in cuprate compounds, Anderson \cite{And} pointed out that the
essential physics of the electrons in the copper-oxide planes
of these materials could be described by the two-dimensional 
Hubbard model.
The model describes tight-binding electrons with a local repulsion 
$U > 0$, as specified by the Hamiltonian
\begin{equation}
 H = \sum_{\bi,\bj} \sum_{\sg} t_{\bi\bj} \,
 c^{\dag}_{\bi\sg} c_{\bj\sg} +
 U \sum_{\bj} n_{\bj\up} n_{\bj\down} \; .
\end{equation}
in standard second quantization notation.
A hopping amplitude $-t$ between nearest neighbors and
an amplitude $-t'$ between next-nearest neighbors on a square
lattice leads to the dispersion relation
$\eps_{\bk} = -2t \, (\cos k_x + \cos k_y) - 4t' \cos k_x \cos k_y$
for single-particle states.

In agreement with the generic phase diagram of the cuprates, the
Hubbard model is an antiferromagnetic insulator at half-filling 
(provided $t'$ is not too big), and 
is expected to become a d-wave superconductor away from half-filling 
in two dimensions already for quite some time.\cite{Sca}
In particular, the exchange of antiferromagnetic spin fluctuations 
has been proposed as a plausible mechanism leading to d-wave 
pairing.\cite{MSV,SLH,BSS} 
It turned out to be very hard to detect superconductivity in the 
Hubbard model by exact numerical computation,\cite{Sca,Dag} as a 
consequence of finite size and/or temperature limitations.

The tendency toward antiferromagnetism and d-wave pairing in the 
two-dimensional Hubbard model is present already at weak coupling. 
However, conventional perturbation theory breaks down in the most
interesting density regime, since competing infrared divergences 
appear as a consequence of Fermi surface nesting and van Hove 
singularities.\cite{Sch87,Dzy87,LMP}
A controlled treatment of these divergences can be achieved by 
a renormalization group (RG) analysis, which takes into account 
the particle-particle and particle-hole channels on equal footing.

A suitable framework for a systematic RG analysis of the 2D Hubbard 
model is provided by the so-called exact or {\em functional} RG.
\cite{Met}
In this approach, fermionic fields in a functional integral 
representation of the model are integrated successively by 
descending step by step in energy scales. 
This can be formulated as an exact hierarchy of flow equations for 
the effective interactions.
The energy scale of the fields, $\Lam$, is the flow parameter.
Several groups have computed the flow of effective two-particle
interactions for the two-dimensional Hubbard model, using various 
versions of the functional RG in one-loop approximation.
\cite{ZS,HM00a,HSFR,KK03a}
Antiferromagnetic and superconducting instabilities were detected
from the flow of the corresponding susceptibilities.

At sufficiently low temperatures, and in particular at $T=0$,
the effective two-particle interaction $\Gam^{\Lam}$ obtained from
a one-loop approximation diverges at a finite scale $\Lam_c > 0$, 
that is, before all fields have been integrated out. 
Hence, one is running into a strong coupling problem in the 
low-energy limit, even in the case of a weak bare interaction.

If the vertex function diverges only in the Cooper channel,
driven by the particle-particle contribution to the flow, the 
strong coupling problem emerging in the low-energy region can be 
controlled by exploiting $\Lam_c$ as a small parameter.\cite{FMRT}
The scale $\Lam_c$ is exponentially small for a small bare 
interaction.
The formation of a superconducting ground state can then be 
described essentially by a BCS mean-field theory with renormalized 
input parameters.
In the two-dimensional Hubbard model, an instability where the
interactions diverge only in the Cooper channel is realized at 
sufficiently small $U$, if the Fermi surface stays away from van 
Hove points.\cite{ZS,HM00a,HSFR,KK03a} 
At van Hove filling the effective interactions diverge also in 
other channels even in the weak coupling limit. Although
superconductivity then has to compete with other instabilities,
the one-loop flow indicates that it is still the leading instability 
for a moderate $t' \neq 0$ and sufficiently small $U$.
\cite{HM00a,HSFR,HM00b}

In principle, spontaneous symmetry breaking can be handled within 
the functional RG framework by adding an infinitesimal symmetry 
breaking term at the beginning of the flow, which is then promoted 
to a finite order parameter at the scale $\Lam_c$.\cite{SHML,GHRM}
So far, this approach has been worked out in practice only for
mean-field models.
Order parameter fluctuations are most conveniently treated by
introducing appropriate bosonic fields, as discussed recently
for antiferromagnetic order in the half-filled Hubbard 
model.\cite{BBW}

In case of competing order parameters, such as antiferromagnetism
and d-wave superconductivity, a full RG treatment of spontaneous 
symmetry breaking and order parameter fluctuations is a rather 
ambitious long-term goal. 
In the present work, we explore a simpler alternative and combine 
the RG with a mean-field (MF) theory of symmetry breaking.
In this RG+MF approach the one-loop flow is stopped at a scale 
$\Lam_{\rm MF}$ that is small compared to the band width, but still 
safely above the scale $\Lam_c$ where the two-particle vertex diverges. 
At this point the vertex has developed already a pronounced 
momentum dependence, reflecting in particular antiferromagnetic 
and superconducting correlations.
The integration over the remaining modes, below $\Lam_{\rm MF}$, is
treated in a mean-field approximation allowing, in particular, 
antiferromagnetic and superconducting order. 
Low-energy fluctuations are thereby neglected. 
The mean-field Hamiltonian is defined on a restricted momentum region 
near the Fermi surface, with $|\eps_{\bk}-\mu| < \Lam_{\rm MF}$, and 
the effective interactions entering the mean-field equations are 
extracted from $\Gam^{\Lam_{\rm MF}}$.
A very short account of some RG+MF results has appeared recently in
Ref.~\onlinecite{MRR}.

Our theory extends previous mean-field treatments of antiferromagnetism
and d-wave superconductivity in two-dimensional Hubbard-type and
t-J models, where the effective interactions were specified by an 
ad hoc ansatz or identified with bare microscopic interactions,
\cite{IDHR,MF,Kyu,YK2} while we compute the effective interactions
by integrating out fluctuations.

In Sec.\ II we will review the functional RG and the structure of
the flow equations for the effective interactions on one-loop level,
focussing on the Wick ordered version used in the present work.
The mean-field equations for antiferromagnetism and 
superconductivity, including a possible coexistence of both, will 
be derived in Sec.\ III.
Results from the combined RG+MF analysis for the 2D Hubbard model
will be presented in Sec.\ IV. In particular, we will analyze which
phases are stabilized for various choices of hopping, interaction and
density, and we will show results for the size and shape (momentum 
dependence) of the relevant order parameters.
We conclude with a summary of the main results in Sec.\ V.


\section{Renormalization group}

All versions of the functional RG for interacting Fermi systems are
variants of Wilson's \cite{WK} momentum shell RG, where fermionic
fields are integrated out by descending successively from the 
modes with highest energy down to the Fermi surface.\cite{Sha}
This type of RG is also the basis for important rigorous work on
two-dimensional Fermi systems.\cite{Sal}
The successive integration of modes can be formulated as an
exact hierarchy of flow equations for effective interactions
(one-particle, two-particle etc.).\cite{Sal,SH}
A similar hierarchy can be obtained by using the interaction
strength instead of a cutoff as flow parameter.\cite{HRAE}
We focus on the Wick-ordered version of the functional RG, which 
we will use in our calculations.

\subsection{Wick ordered flow equations}

Before turning to approximations, let us first sketch the
structure and derivation of the exact flow equations (for 
details, see Salmhofer \cite{Sal} and Ref.\ \onlinecite{HM00a}).
The flow parameter $\Lam$ is introduced as an infrared cutoff
for the bare propagator, such that contributions from states
with momenta obeying $|\eps_{\bk} - \mu| < \Lam$ are suppressed.
All Green functions of the interacting system will then depend 
on $\Lam$, and the true theory is recovered only in the limit 
$\Lam \to 0$.
The RG equations are most conveniently obtained from the
effective potential $\cV^{\Lam}$, which is the generating 
functional for connected Green functions with bare propagators 
amputated from the external legs.
Taking a $\Lam$-derivative one obtains an exact functional flow
equation for this quantity.
Expanding $\cV^{\Lam}$ on both sides of the flow equation in 
powers of the fermionic fields (i.e.\ Grassmann variables),
and comparing coefficients, one obtains the so-called Polchinski 
equations \cite{Pol} for the effective m-body interactions.
These equations were used by Zanchi and Schulz \cite{ZS} in
their RG analysis of the two-dimensional Hubbard model.
An alternative expansion in terms of Wick ordered monomials
of fermion fields yields flow equations for the corresponding 
m-body interactions $V^{\Lam}_m$ with a particularly convenient
structure (see Fig.\ 1).\cite{Sal}
\begin{figure}[ht]
\centering
\includegraphics[width = 10cm]{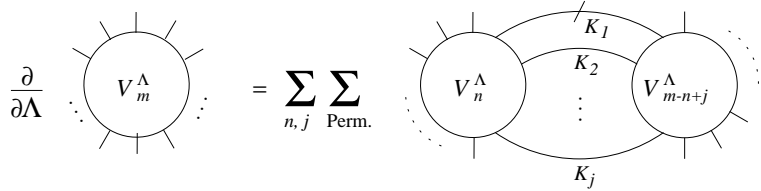}
\caption{Diagrammatic representation of the flow equation for 
 $V^{\Lam}_m$ in the Wick ordered version of the functional RG. 
 The line with a slash corresponds to 
 $\partial G_0^{<\Lam}/\partial\Lam$,
 the others to $G_0^{<\Lam}$; all possible pairings leaving $m$ 
 incoming and $m$ outgoing external legs have to be summed.}
\label{fig1}
\end{figure}
The flow of $V^{\Lam}_m$ is given as a bilinear form of other
n-body interactions (at the same scale $\Lam$), which are 
connected by lines corresponding to the propagator
\begin{equation}
 G_0^{<\Lam}(k_0,\bk) = 
 \frac{\Theta(\Lam - |\xi_{\bk}|)}{ik_0 - \xi_{\bk}} \; ,
\end{equation}
where $\xi_{\bk} = \eps_{\bk} - \mu$, and one line corresponding 
to $\partial_{\Lam} G_0^{<\Lam}(k_0,\bk)$.
Note that the support of $G_0^{<\Lam}(k_0,\bk)$ is restricted to 
momenta with $|\xi_{\bk}|$ {\em below}\/ the scale $\Lam$.
These soft mode propagators come into play via the Wick 
ordering, which is given by (scale dependent) contractions with 
$G_0^{<\Lam}(k_0,\bk)$.
For $\Lam \to 0$, the effective interactions remain unaffected 
by the Wick ordering, since $G_0^{<\Lam}(k)$ vanishes in that
limit.
For small $\Lam$, the momentum integrals on the right hand side
of the flow equation are restricted to momenta close to the
Fermi surface (see Fig.\ 2).
\begin{figure}[ht]
\centering
\includegraphics[clip=true,width=6cm]{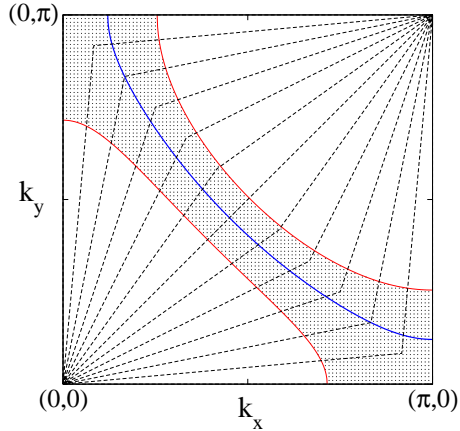}
\caption{(Color online) 
 A typical Fermi surface (bold line) and the support of 
 the propagator $G_0^{<\Lam}(k_0,\bk)$ (dotted region) in the first
 quadrant of the Brillouin zone. 
 The dashed lines mark the boundaries of the ''patches'' used for
 the discretization of the momentum dependence of the effective
 two-particle interaction.}
\label{fig2}
\end{figure}

With the initial condition $\cV^{\Lam_0} = \mbox{bare interaction}$,
where $\Lam_0 = \max |\xi_{\bk}|$, the above flow equations 
determine the exact flow of the effective interactions as $\Lam$ 
sweeps over the entire energy range from the band edges down 
to the Fermi level.
Since the Wick ordered flow below scale $\Lam$ involves only 
low-energy states with $|\xi_{\bk}| < \Lam$, it yields a continuous 
sequence of effective low-energy models with effective interactions 
acting on a restricted momentum space.

\subsection{One-loop flow}

To detect instabilities of the system in the weak-coupling limit,
it is sufficient to truncate the infinite hierarchy of flow 
equations described by Fig.\ 1 at second order in the effective
two-particle interaction $V^{\Lam}_2$.
Contributions to the flow of $V^{\Lam}_2$ involving effective
three-particle interactions and higher $m$-body terms are at 
least of third order in $V^{\Lam}_2$.
The leading contribution from the effective one-particle 
interaction $V_1^{\Lam}$ can be absorbed by a simple shift of 
the chemical potential. The remaining influence of $V_1^{\Lam}$ 
on the flow of $V^{\Lam}_2$ is of third order an can thus be
neglected.

The flow of the effective two-particle interaction $V^{\Lam}_2$ 
is thus given by one-loop terms involving only $V^{\Lam}_2$, 
and no other $m$-body terms. 
In the following, we write $\Gam^{\Lam}$ instead of $V^{\Lam}_2$ 
for the two-particle interaction.
Putting arrows on the lines to distinguish creation and 
annihilation operators, one obtains the diagrammatic representation
of the flow equation shown in Fig.\ 3.
\begin{figure}[ht]
\centering
\includegraphics[width = 10cm]{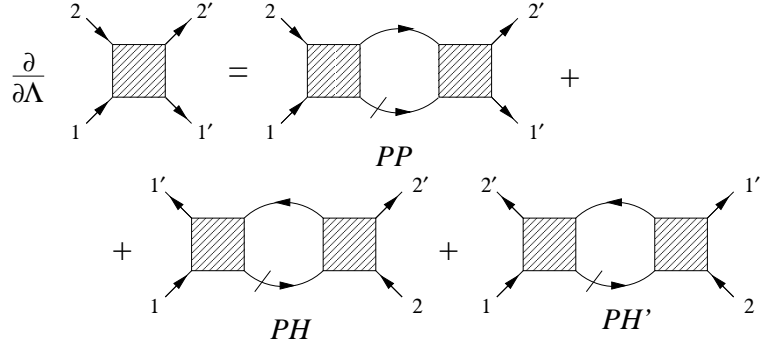}
\caption{Flow equation for the effective two-particle interaction
 $\Gam^{\Lam}$ in one-loop approximation with the particle-particle
 channel (PP) and the two particle-hole channels (PH and PH').}
\label{fig3}
\end{figure}
To write the momentum, energy and spin dependences of 
$\Gam^{\Lam}$, we collect $k=(k_0,\bk)$ and $\sg$ in a 
single variable $K$. In $\Gam^{\Lam}(K'_1,K'_2;K_1,K_2)$, 
the variables $K_1$ and $K_2$ refer to incoming, the variables  
$K'_1$ and $K'_2$ to outgoing particles.
With this notation, the explicit one-loop flow equation for 
$\Gam^{\Lam}$ reads \cite{HM00a}
\begin{equation} \label{1loop}
 \renewcommand{\arraystretch}{1.2}
 \begin{split}
 \frac{\partial}{\partial\Lam} 
 \Gam^{\Lam}(K'_1&,K'_2; K_1,K_2) = \\
 \, & \frac{1}{\beta L} \sum_{K,K'}
 \frac{\partial}{\partial\Lam} \left[
  G_0^{<\Lam}(K) \, G_0^{<\Lam}(K') \right] \\
 & \times  \Big[ 
  \frac{1}{2} 
    \Gam^{\Lam}(K'_1,K'_2;K,K') \, \Gam^{\Lam}(K,K';K_1,K_2)
 \\ & \quad
  - \Gam^{\Lam}(K'_1,K;K_1,K') \, \Gam^{\Lam}(K',K'_2;K,K_2)
 \\ & \quad
  + \Gam^{\Lam}(K'_2,K;K_1,K') \, \Gam^{\Lam}(K',K'_1;K,K_2)
   \Big] \; ,
 \end{split}
\end{equation}
where $L$ is the number of lattice sites,
and $\beta$ the inverse temperature.
The three terms on the right hand side are the contributions
from the particle-particle channel (PP) and the two particle-hole 
channels (PH and PH').
For translation invariant systems momentum 
conservation implies that $\Gam^{\Lam}(K'_1,K'_2;K_1,K_2) \neq 0$
only if $k_1 + k_2 = k'_1 + k'_2$, so that the sum over $k$ and 
$k'$ in (\ref{1loop}) is reduced to a single energy-momentum sum.

For a spin-rotation invariant system the spin structure of the 
two-particle interaction can be written as
\begin{equation}
 \Gam^{\Lam}(K'_1,K'_2;K_1,K_2) =
 \Gam^{\Lam}_s(k'_1,k'_2;k_1,k_2) \, 
 S_{\sg'_1,\sg'_2;\sg_1,\sg_2} +
 \Gam^{\Lam}_t(k'_1,k'_2;k_1,k_2) \, 
 T_{\sg'_1,\sg'_2;\sg_1,\sg_2} \; ,
\end{equation}
where 
\begin{equation}
\begin{split}  
 S_{\sg'_1,\sg'_2;\sg_1,\sg_2} = 
 \textstyle \frac{1}{2} \, \left(
 \delta_{\sg_1\sg'_1}\delta_{\sg_2\sg'_2} -
 \delta_{\sg_1\sg'_2}\delta_{\sg_2\sg'_1} \right) 
 \\
 T_{\sg'_1,\sg'_2;\sg_1,\sg_2} = 
 \textstyle \frac{1}{2} \, \left(
 \delta_{\sg_1\sg'_1}\delta_{\sg_2\sg'_2} +
 \delta_{\sg_1\sg'_2}\delta_{\sg_2\sg'_1} \right)
\end{split}
\end{equation}
are the projection operators on singlet and triplet states
in a two-particle spin space, respectively.
The antisymmetry of $\Gam^{\Lam}$ with respect to 
$K_1 \leftrightarrow K_2$ or $K'_1 \leftrightarrow K'_2$
implies that $\Gam^{\Lam}_s$ is symmetric and $\Gam^{\Lam}_t$
antisymmetric under exchange of the variables $k_1$ and $k_2$
or $k'_1$ and $k'_2$.
Inserting the above decomposition in singlet and triplet terms,
the spin sums can be easily carried out and one obtains a coupled
set of equations for $\Gam^{\Lam}_s$ and $\Gam^{\Lam}_t$.

It is clearly impossible to solve the flow equations with the full
energy and momentum dependence of the two-particle interaction, since
$\Gam^{\Lam}$ has three independent energy and momentum variables.
The problem can however be simplified by ignoring dependences 
which are {\em irrelevant} (in the RG sense) in the low energy limit, 
namely the energy dependence and the momentum dependence normal to 
the Fermi surface.\cite{Sha}
Hence, we compute the flow of the two-particle interaction at zero 
energy and with at least three momenta on the Fermi surface (the 
fourth being determined by momentum conservation). 
On the right hand side of the flow equation we approximate the 
interaction by its zero energy value with three momenta projected 
on the Fermi surface (if not already there).
Note that this projection procedure is exact for the initial
two-particle interaction, since the Hubbard interaction is 
momentum and energy independent.
The remaining tangential momentum dependence is discretized.
The momentum-dependence of the two-particle vertex is thus 
approximated by a step function which is constant on ''patches'' 
(sectors) in the Brillouin zone.\cite{ZS,HM00a,HSFR}
Here we use patches defined by straight lines connecting the
points $(0,0)$ and $(\pi,\pi)$ to the magnetic Brillouin zone
boundary (also known as ''umklapp surface''), see Fig.~2.

Neglecting the energy dependence of $\Gam^{\Lam}$, the
Matsubara sum in the flow equation can be done analytically.
Due to the sharp momentum cutoff the momentum integral can be
easily reduced to a one-dimensional integral. The latter, and
the integration of the flow, has to be performed numerically.

We will stop the flow at a scale $\Lam_{\rm MF}$ well above the scale
$\Lam_c$ at which $\Gam^{\Lam}$ diverges.
The remaining degrees of freedom will be treated in a mean-field
approximation to be described in the next section.
The output of the flow, $\Gam^{\Lam_{\rm MF}}$, is the input 
interaction for the mean-field theory.


\section{Mean-field theory}

The divergence of the effective two-particle interaction 
$\Gam^{\Lam}$ as a function of decreasing $\Lam$ signals an 
instability leading to an ordered phase via spontaneous 
symmetry breaking.\cite{fn1}
The simplest way to treat spontaneous symmetry breaking is 
provided by a mean-field approximation. 
For a specific type of order, only interactions with a very
restricted choice of momenta are picked up by the mean-field
theory. 
For example, in the BCS mean-field theory of superconductivity,
the only relevant interaction processes involve particles 
with vanishing total momentum ($\bk_1 + \bk_2 = 0$).
Vice versa, in the thermodynamic limit the mean-field 
approximation provides the exact solution of a {\em reduced}\/ 
Hamiltonian, in which the interaction is restricted to the 
relevant momenta from the beginning.

\subsection{Reduced Hamiltonian}

Here we specify our reduced Hamiltonian, and link the
interaction terms to the effective two-particle interaction
obtained from the RG. In addition to interactions driving
antiferromagnetism and superconductivity, we will also
include several other interaction terms, since they do not
lead to major complications, make the general structure of
the theory more transparent, and may be useful for a more
general analysis in the future.

The reduced Hamiltonian has the form
\begin{equation}
 H^{\rm red} = H_0 + H_{\rm I}^{\rm red} \; ,
\end{equation}
where 
$H_0 = \sum_{\bk,\sg} \eps_{\bk} \, n_{\bk\sg}$ 
is the kinetic energy and $H_{\rm I}^{\rm red}$ the
reduced interaction. The latter contains four terms,
\begin{equation}
 H_{\rm I}^{\rm red} = 
 H_{\rm I}^{n,0} + H_{\rm I}^{n,\pi} + 
 H_{\rm I}^{p,0} + H_{\rm I}^{p,\pi} \; ,
\end{equation}
which we now specify one by one.

The first term is a density-density interaction with zero 
momentum transfer (forward scattering),
\begin{equation}
 H_{\rm I}^{n,0} =
 \frac{1}{2L} \sum_{\bk,\bk'} \sum_{\sg,\sg'} 
 F_{\bk\bk'}^{\sg\sg'} \, n_{\bk\sg} \, n_{\bk'\sg'} \; .
\end{equation}
This term captures Fermi liquid interaction effects, and can 
lead to charge or spin density instabilities such as 
phase separation, ferromagnetism, or a d-wave Pomeranchuk 
instability \cite{HM00b,YK1}.

The second term is a density-density interaction with 
momentum transfer $\bQ = (\pi,\pi)$,
\begin{equation}
 H_{\rm I}^{n,\pi} =
 \frac{1}{2L} \sum_{\bk,\bk'} \sum_{\sg,\sg'} 
 U_{\bk\bk'}^{\sg\sg'} \, 
 n_{\bk\sg}^{\pi} \, n_{\bk'\sg'}^{\pi} \; ,
\end{equation}
with 
$n_{\bk\sg}^{\pi} = 
 a_{\bk\sg}^{\dag} a_{\bk+\bQ,\sg}^{\phantom\dag}$.
Note that $\bk+\bQ = \bk-\bQ$.
This term drives charge or spin density wave instabilities with a 
wave vector $\bQ$ and arbitrary form factors (s-wave, d-wave etc.), 
including, in particular, antiferromagnetic order.

The third term is a singlet pairing interaction between particles 
with total momentum zero (Cooper channel),
\begin{equation}
 H_{\rm I}^{p,0} =
 \frac{1}{L} \sum_{\bk,\bk'} V_{\bk\bk'} \, 
 p_{\bk}^{\dag} \, p_{\bk'}^{\phantom\dag} \; ,
\end{equation}
with
$p_{\bk} = a_{-\bk\down} a_{\bk\up}$ and
$p_{\bk}^{\dag}$ its hermitian conjugate.
There is no factor $1/2$ here, since spin variables are fixed.
This term drives spin-singlet superconductivity.

Finally, we include a triplet pairing interaction between
particles with a total momentum $\bQ$,
\begin{equation}
 H_{\rm I}^{p,\pi} =
 \frac{1}{L} \sum_{\bk,\bk'} V^{\pi}_{\bk\bk'} \, 
 p_{\bk}^{\pi \, \dag} \, p_{\bk'}^{\pi} \; ,
\end{equation}
with
$p_{\bk}^{\pi} = a_{\bQ-\bk,\down} a_{\bk\up}$ and
$p_{\bk}^{\pi \, \dag}$ its hermitian conjugate.
In case of coexistence of antiferromagnetism and 
superconducitivity, this term leads to a condensate 
of pairs with total momentum $\bQ$.\cite{PF,MF,Kyu}

We use the mean-field theory as an approximate treatment
of the fermionic degrees of freedom below the energy scale
$\Lam_{\rm MF}$. Hence, all momenta in the above terms are 
restricted to a shell around the Fermi surface given by
$|\eps_{\bk} - \mu| < \Lam_{\rm MF}$.
For the dispersion relation $\eps_{\bk}$ in $H_0$, we simply 
use the bare one, that is, we do not keep track of self-energy
contributions which may renormalize $\eps_{\bk}$. 
Self-energy corrections beyond those which can be absorbed
in a shift of $\mu$ appear only at second order in the 
interactions, and may therefore be neglected at weak 
coupling.

The coupling functions in $H_{\rm I}^{\rm red}$ are extracted
from the effective two-particle interaction $\Gam^{\Lam_{\rm MF}}$.
Denoting the static (frequency independent) two-particle 
interaction with generic momenta by 
$\Gam^{\Lam_{\rm MF}}
 (\bk'_1\sg'_1,\bk'_2\sg'_2;\bk_1\sg_1,\bk_2\sg_2)$,
we have
\begin{eqnarray}
 F_{\bk\bk'}^{\sg\sg'} &=& 
 \frac{1}{2} \, \Gam^{\Lam_{\rm MF}}(\bk\sg,\bk'\sg';\bk\sg,\bk'\sg') 
 \; , \\[2mm]
 U_{\bk\bk'}^{\sg\sg'} &=& 
 \frac{1}{2} \, 
 \Gam^{\Lam_{\rm MF}}(\bk\sg,\bk'\sg';\bk+\bQ\,\sg,\bk'-\bQ\,\sg')  
 \; , \\[2mm]
 V_{\bk\bk'} &=& \frac{1}{2} \, 
 \Gam^{\Lam_{\rm MF}}_s(\bk,-\bk;\bk',-\bk') \; , \\[2mm]
 V^{\pi}_{\bk\bk'} &=& \frac{1}{2} \, 
 \Gam^{\Lam_{\rm MF}}_t(\bk,\bQ-\bk;\bk',\bQ-\bk') \; .
\end{eqnarray}
All these coupling functions are real-valued.
Note that there is an overlap between the different coupling
functions above for special choices of momenta. For example,
for $\bk_1 = \bk'_1 = -\bk_2 = -\bk'_2$ the vertex belongs
to the Cooper and forward scattering channels simultaneously.
However, these special momentum sets have zero measure in the
thermodynamic limit.

\subsection{Mean-field decoupling}

The mean-field theory involves the following mean fields
\begin{eqnarray}
 D_{\bk\sg} &=& \frac{1}{L} \sum_{\bk'\sg'} 
 F_{\bk\bk'}^{\sg\sg'} \, \bra n_{\bk'\sg'} \ket \\
 D_{\bk\sg}^{\pi} &=& \frac{1}{L} \sum_{\bk'\sg'} 
 U_{\bk\bk'}^{\sg\sg'} \, \bra n_{\bk'\sg'}^{\pi} \ket \\
 \Delta_{\bk} &=& \frac{1}{L} \sum_{\bk'} 
 V_{\bk\bk'} \, \bra p_{\bk'} \ket \\
 \Delta_{\bk}^{\pi} &=& \frac{1}{L} \sum_{\bk'} 
 V^{\pi}_{\bk\bk'} \, \bra p_{\bk'}^{\pi} \ket \; .
\end{eqnarray}
The density field $D_{\bk\sg}$ is real. 
The density field $D^{\pi}_{\bk\sg}$ can be complex and
obeys the relation
$D^{\pi}_{\bk+\bQ,\sg} = D^{\pi *}_{\bk\sg}$,
which follows directly from 
$\bra n^{\pi}_{\bk+\bQ,\sg} \ket = \bra n^{\pi}_{\bk\sg} \ket^*$
and
$U^{\sg\sg'}_{\bk+\bQ,\bk'+\bQ} = U^{\sg\sg'}_{\bk\bk'} \,$.
The pairing fields $\Delta_{\bk}$ and $\Delta^{\pi}_{\bk}$
are generally complex.

The operator products in the interaction terms have the form
$b'b$, where $b$ and $b'$ are products of two elementary Fermi 
operators. In the mean-field decoupling, one approximates
$b'b \approx \bra b' \ket \, b + \bra b \ket \, b' - 
 \bra b' \ket \, \bra b \ket$. 
The fluctuation term $(b' - \bra b' \ket) \, (b - \bra b \ket)$ 
is thereby neglected.
This yields
\begin{eqnarray}
 H_{\rm I}^{n,0} &\approx&
 \sum_{\bk,\sg} D_{\bk\sg} \, n_{\bk\sg}  -
 \frac{1}{2} \sum_{\bk,\sg} D_{\bk\sg} \bra n_{\bk\sg} \ket
 \; , \\[2mm]
 H_{\rm I}^{n,\pi} &\approx&
 \sum_{\bk,\sg} D^{\pi}_{\bk\sg} \, n^{\pi}_{\bk\sg}  -
 \frac{1}{2} \sum_{\bk,\sg} D^{\pi}_{\bk\sg} 
 \bra n^{\pi}_{\bk\sg} \ket \; , \\[2mm]
 H_{\rm I}^{p,0} &\approx&
 \sum_{\bk} \Delta^*_{\bk} \, p_{\bk} +
 \sum_{\bk} \Delta_{\bk} \, p^{\dag}_{\bk} -
 \sum_{\bk} \Delta^*_{\bk} \, \bra p_{\bk} \ket \; , \\[2mm]
 H_{\rm I}^{p,\pi} &\approx&
 \sum_{\bk} \Delta^{\pi *}_{\bk} \, p^{\pi}_{\bk} +
 \sum_{\bk} \Delta^{\pi}_{\bk} \, p^{\pi \, \dag}_{\bk} -
 \sum_{\bk} \Delta^{\pi *}_{\bk} \, \bra p^{\pi}_{\bk} \ket
 \; .
\end{eqnarray}
For the above reduced interactions this approximation is
exact in the thermodynamic limit. This can be seen quite
easily by estimating the size of Feynman diagrams with a 
reduced interaction or, alternatively, by using a Hubbard
Stratonovich decoupling in a functional integral representation.
\cite{Muh} 

The Hamiltonian is now quadratic in the fermions and can
therefore be diagonalized. 
We define $K^{\rm MF} = H^{\rm MF} - \mu N$, where
$H^{\rm MF}$ is
the reduced Hamiltonian in mean-field approximation, and
$N = \sum_{\bk,\sg} n_{\bk\sg}$ is the total particle number
operator.
Introducing the Nambu spinor 
\begin{equation}
 \ba_{\bk} = (a_{\bk\up}, a^{\dag}_{-\bk\down}, 
 a_{\bk+\bQ,\up}, a^{\dag}_{-\bk-\bQ,\down}) \; ,
\end{equation}
the mean-field Hamiltonian can be written as
\begin{equation}
 K^{\rm MF} = 
 \sum_{\bk} \ba^{\dag}_{\bk} \cM_{\bk} \, \ba_{\bk} + K^c 
 \; ,
\end{equation}
where $K^c$ collects all c-number terms, that is, terms 
without operators, and the quadratic form with the matrix
$\cM_{\bk}$ all quadratic terms. 
Here the momentum sum extends only over momenta in the 
{\em magnetic}\/ Brillouin zone.
The matrix $\cM_{\bk}$ is given by
\begin{equation}
 \cM_{\bk} = \left(
 \begin{array}{cccc}
  \xi_{\bk} + D_{\bk\up} & \Delta_{\bk} & 
  D^{\pi}_{\bk\up} & \Delta^{\pi}_{\bk} \\
  \Delta^*_{\bk} & - \xi_{-\bk} - D_{-\bk\down} &
  \Delta^{\pi *}_{\bk+\bQ} & - D^{\pi}_{-\bk-\bQ,\down} \\
  D^{\pi}_{\bk+\bQ,\up} & \Delta^{\pi}_{\bk+\bQ} &
  \xi_{\bk+\bQ} + D_{\bk+\bQ,\up} & \Delta_{\bk+\bQ} \\
  \Delta^{\pi *}_{\bk} & - D^{\pi}_{-\bk\down} &
  \Delta^*_{\bk+\bQ} & - \xi_{-\bk-\bQ} - D_{-\bk-\bQ,\down} 
 \end{array} \right) \; .
\end{equation}
Note that $\cM_{\bk}$ is hermitean since
$D^{\pi}_{\bk+\bQ,\sg} = D^{\pi *}_{\bk\sg}$.
The momenta of the mean fields in $\cM_{\bk}$ are 
restricted by the cutoff:
$D_{\bk\sg}$ and $\Delta_{\bk}$ are restricted by the
condition $|\xi_{\bk}| < \Lam_{\rm MF}$, while
$D^{\pi}_{\bk}$ and $\Delta^{\pi}_{\bk}$ contribute only 
if the two conditions $|\xi_{\bk}| < \Lam_{\rm MF}$ and 
$|\xi_{\bk+\bQ}| < \Lam_{\rm MF}$ are satisfied simultaneously.
The matrix $\cM_{\bk}$ has a $2\times 2$ block structure
\begin{equation} \label{block}
 \cM_{\bk} = \left(
 \begin{array}{cc}
 {\cal M}^0_{\bk} & {\cal M}^{\pi}_{\bk} \\
 {\cal M}^{\pi}_{\bk+\bQ} & {\cal M}^0_{\bk+\bQ}
\end{array} \right) \; ,
\end{equation}
where ${\cal M}^0_{\bk}$ and ${\cal M}^{\pi}_{\bk}$ are 
$2\times 2$ matrices.

The c-number term, $K^c$ contains a contribution from 
commutators, $\sum_{\bk} (\xi_{\bk} + D_{\bk\down})$,
in addition to the c-number terms originating from the
mean-field decoupling of the interaction terms.

The matrix $\cM_{\bk}$ can be diagonalized by a unitary 
transformation $\cU_{\bk}$:
\begin{equation}
 \sum_{\bk} \ba^{\dag}_{\bk} \cM_{\bk} \, \ba_{\bk} =
 \sum_{\bk} \tba^{\dag}_{\bk} \tilde\cM_{\bk} \, \tba_{\bk}
 \; ,
\end{equation}
where
$\tba_{\bk} = \cU_{\bk} \, \ba_{\bk}$, 
$\tba^{\dag}_{\bk} = \ba^{\dag}_{\bk} \, \cU^{\dag}_{\bk}$, and
$\tilde\cM_{\bk} = \cU_{\bk} \, \cM_{\bk} \, \cU^{\dag}_{\bk}$ 
is a diagonal matrix 
\begin{equation}
 \tilde\cM_{\bk} = 
 {\rm diag}(E^1_{\bk},E^2_{\bk},E^3_{\bk},E^4_{\bk})
\end{equation}
with the eigenvalues $E^{\lam}_{\bk}$, $\lam = 1,2,3,4$,
as entries. 
The transformed creation and annihilation operators obey
fermionic anticommutation relations,
$\{ \tilde a^{\lam}_{\bk}, \tilde a^{\lam'\dag}_{\bk'} \} =
 \delta_{\lam\lam'} \, \delta_{\bk\bk'} \,$, since $\cU_{\bk}$
is unitary.
The block structure (\ref{block}) implies that the matrices
$\cM_{\bk}$ and $\cM_{\bk+\bQ}$ have the same eigenvalues,
which can thus be labelled such that  
$E^{\lam}_{\bk+\bQ} = E^{\lam}_{\bk}$.

\subsection{General gap equations}

The selfconsistency or gap equations can be derived by 
minimizing the grand canonical potential
\begin{equation} \label{Omega}
 \Omega = K^c - T \sum_{\bk,\lam} 
 \ln \left( 1 + e^{-\beta \, E^{\lam}_{\bk}} \right)
\end{equation}
with respect to the expectation values $\bra n_{\bk\sg} \ket$,
$\bra n^{\pi}_{\bk\sg} \ket$, $\bra p_{\bk} \ket$,
$\bra p^{\pi}_{\bk} \ket$, or complex conjugates.
As shown in Appendix A, the gap equations can be expressed
in terms of minors of the matrices
\begin{equation}
 \cM^{\lam}_{\bk} = \cM_{\bk} - E^{\lam}_{\bk} {\bf 1} \; .
\end{equation}
The minor $\det(M^{\lam,jj'}_{\bk})$ is the determinant of the matrix 
obtained from $\cM^{\lam}_{\bk}$ by deleting the $j$-th row and the 
$j'$-th column.
Minimizing with respect to $\bra n_{\bk\sg} \ket$ yields
\begin{equation} \label{gapeq1}
 D_{\bk\sg} = \frac{1}{L} \sum_{\bk'} \sum_{\lam} 
 \frac{F^{\sg\up}_{\bk\bk'} \,
       \det \big( M^{\lam,11}_{\bk'} \big) -
       F^{\sg\down}_{\bk,-\bk'} \,
       \det \big( M^{\lam,22}_{\bk'} \big)}
 {\sum_l \det \big( M^{\lam,ll}_{\bk'} \big)} 
 \, f(E^{\lam}_{\bk'}) +
 \frac{1}{L} \sum_{\bk'} F^{\sg\down}_{\bk\bk'} \; ,
\end{equation}
where $f$ is the Fermi function. 
The last term in (\ref{gapeq1}) stems from the contribution 
$\sum_{\bk} D_{\bk\down}$ to $K^c$ mentioned above.
Minimizing with respect to $\bra n^{\pi}_{\bk\sg} \ket$ yields
\begin{equation} \label{gapeq2}
 D^{\pi}_{\bk\sg} = \frac{1}{L} \sum_{\bk'} \sum_{\lam} 
 \frac{U^{\sg\up}_{\bk\bk'} \,
       \det \big( M^{\lam,13}_{\bk'} \big) -
       U^{\sg\down}_{\bk,-\bk'} \,
       \det \big( M^{\lam,42}_{\bk'} \big)}
 {\sum_l \det \big( M^{\lam,ll}_{\bk'} \big)} 
 \, f(E^{\lam}_{\bk'}) \; .
\end{equation}
Minimization with respect to $\bra p_{\bk} \ket^*$ yields
\begin{equation} \label{gapeq3}
 \Delta_{\bk} = 
 - \frac{1}{L} \sum_{\bk'} V_{\bk\bk'} \sum_{\lam}
 \frac{\det \big( M^{\lam,21}_{\bk'} \big)}
 {\sum_l \det \big( M^{\lam,ll}_{\bk'} \big)} 
 \, f(E^{\lam}_{\bk'}) \; ,
\end{equation}
while minimizing with respect to $\bra p_{\bk} \ket$ would
just yield the complex conjugate of the above equation.
Minimization with respect to $\bra p^{\pi}_{\bk} \ket^*$
yields 
\begin{equation} \label{gapeq4}
 \Delta^{\pi}_{\bk} = 
 - \frac{1}{L} \sum_{\bk'} V^{\pi}_{\bk\bk'} \sum_{\lam}
 \frac{\det \big( M^{\lam,41}_{\bk'} \big)}
 {\sum_l \det \big( M^{\lam,ll}_{\bk'} \big)} 
 \, f(E^{\lam}_{\bk'}) \; ,
\end{equation}
and minimization with respect to $\bra p^{\pi}_{\bk} \ket$
again the complex conjugate.
Note that the momentum sums in the above gap equations 
extend over the full (not just magnetic) Brillouin zone 
within the limits imposed by the cutoff.

The denominator in the gap equations can be expressed
in terms of the eigenvalues as (see Appendix A)
\begin{equation} \label{denom}
 \sum_l \det \big( M^{\lam,ll}_{\bk} \big) =
 \prod_{\lam' \neq \lam} 
 ( E^{\lam'}_{\bk} - E^{\lam}_{\bk} ) \; .
\end{equation}
The electron density $n = - L^{-1} \partial\Om/\partial\mu$ 
can be written as
\begin{equation} \label{density}
 n = 1 + \frac{1}{L} \sum_{\bk'} \sum_{\lam} 
 \frac{\det \big( M^{\lam,11}_{\bk'} \big) -
       \det \big( M^{\lam,22}_{\bk'} \big)}
 {\sum_l \det \big( M^{\lam,ll}_{\bk'} \big)} 
 \, f(E^{\lam}_{\bk'}) \; . 
\end{equation}
The derivation of this relation is almost identical to the 
one of the gap equation for $D_{\bk\sg}$.

For a numerical evaluation it is sometimes convenient to rewrite the 
ratios of determinants in the above expressions in terms of the 
eigenvectors of $\cM_{\bk}$.\cite{Rei}

\subsection{Gap equations for antiferromagnetism and superconductivity}

We are mainly interested in the interplay of antiferromagnetism
and superconductivity in the two-dimensional Hubbard model.
In the numerical solution of the mean-field equations we will
therefore ignore $D_{\bk\sg}$. Furthermore, we will restrict 
$D^{\pi}_{\bk\sg}$ to the spin density wave channel, that is,
$D^{\pi}_{\bk\up} = - D^{\pi}_{\bk\down} \equiv A_{\bk}$.
The mean-field $A_{\bk}$ is related to the staggered magnetisation
$m_{\bk} = n^{\pi}_{\bk\up} - n^{\pi}_{\bk\down}$ by
\begin{equation}
 A_{\bk} = \frac{1}{L} \sum_{\bk'} U^S_{\bk\bk'} \, \bra m_{\bk'} \ket \; ,
\end{equation}
where $U^S_{\bk\bk'} = 
\frac{1}{2} (U^{\sg\sg}_{\bk\bk'} - U^{\sg,-\sg}_{\bk\bk'})$.
For a conventional commensurate antiferromagnet, $A_{\bk}$ is real.
The matrix $\cM_{\bk}$ then assumes the simpler structure
\begin{equation} \label{Mafsc}
 \cM_{\bk} = \left(
 \begin{array}{cccc}
  \xi_{\bk} & \Delta_{\bk} & 
  A_{\bk} & \Delta^{\pi}_{\bk} \\
  \Delta^*_{\bk} & - \xi_{\bk} &
  \Delta^{\pi *}_{\bk+\bQ} & A_{\bk} \\
  A_{\bk} & \Delta^{\pi}_{\bk+\bQ} &
  \xi_{\bk+\bQ} & \Delta_{\bk+\bQ} \\
  \Delta^{\pi *}_{\bk} & A_{\bk} &
  \Delta^*_{\bk+\bQ} & - \xi_{\bk+\bQ} 
 \end{array} \right) \; .
\end{equation}
We have used reflection symmetry to replace $\xi_{-\bk}$ by
$\xi_{\bk}$ etc. 

In principle, the coexistence of antiferromagnetic order and 
superconductivity leads automatically to a finite $\pi$-pairing
field.\cite{PF,MF,Kyu}
To see this, consider the right hand side of the gap equation 
(\ref{gapeq4}) for $\Delta^{\pi}_{\bk}$. The relevant minor in the 
numerator can be written as
\begin{equation} \label{detMl41}
 \det \big( M^{\lam,41}_{\bk} \big) = 
 \Delta_{\bk} A_{\bk} (E^{\lam}_{\bk} - \xi_{\bk+\bQ}) +
 \Delta_{\bk+\bQ} A_{\bk} (E^{\lam}_{\bk} + \xi_{\bk})
\end{equation} 
in the limit $\Delta^{\pi}_{\bk} \to 0$. This is nonzero if both 
$A_{\bk} \neq 0$ and $\Delta_{\bk} \neq 0$. In other words,
the coexistence of antiferromagnetic order and superconductivity
generates a finite expectation value $\bra p^{\pi}_{\bk} \ket$.
For a nonzero interaction in the $\pi$-pairing channel,
$V^{\pi}_{\bk\bk'}$, this leads to a finite $\pi$-pairing field
$\Delta^{\pi}_{\bk}$.
Since both $A_{\bk}$ and $\Delta_{\bk}$ are odd under spin flips,
the spin flip symmetry of the expectation value 
$\bra p^{\pi}_{\bk} \ket$ generated by $A_{\bk}$ and 
$\Delta^{\pi}_{\bk}$ is even, corresponding to triplet 
pairing.
This is why the interaction $V^{\pi}_{\bk\bk'}$ was extracted
from the triplet component of the vertex in Eq.~(15).
In case of coexistence of superconductivity with a charge 
(instead of spin) density wave, the generated $\pi$-pairing
would be odd under spin flips, and thus couples to the singlet
vertex.
Combined with the antisymmetry under particle exchange the
spin flip symmetry of the $\pi$-pairing field yields the relation
$\Delta^{\pi}_{\bk} = - \Delta^{\pi}_{\bQ-\bk}$.

At and near half-filling, the dominant ordering tendencies of the
two-dimensional Hubbard model are antiferromagnetism and singlet
superconductivity. 
Under point group transformations the antiferromagnetic order 
parameter $A_{\bk}$ has s-wave symmetry, while the superconducting 
gap $\Delta_{\bk}$ has d$_{x^2-y^2}$-symmetry. 
From Eq.~(\ref{detMl41}) one can thus deduce that 
$\Delta^{\pi}_{\bk}$ also has d$_{x^2-y^2}$-symmetry. 
Note that the latter is not in conflict with the triplet spin 
symmetry of the $\pi$-pair, since the total momentum of the pair 
is $\bQ = (\pi,\pi)$, not zero. 
The effective interaction $\Gam^{\Lam}$ of the Hubbard model 
turns out to be very small in the triplet $\pi$-pairing channel, 
and its d-wave component is repulsive. Hence the $\pi$-pairing
field $\Delta^{\pi}_{\bk}$ is also very small compared to 
the other order parameters, such that its feedback into the gap
equations for $A_{\bk}$ and $\Delta_{\bk}$ can be neglected.



\section{Results for the two-dimensional Hubbard model}

In this section we present results obtained from the renormalized
mean-field theory on antiferromagnetic order and superconductivity 
in the repulsive two-dimensional Hubbard model. We will first 
discuss antiferromagnetic states in the absence of superconductivity,
then d-wave superconductivity in the absence of antiferromagnetic
order, and finally the full theory allowing for coexistence.
Superconductivity in the attractive Hubbard model is discussed
briefly in Appendix B. 
All results will be presented for $t=1$, that is, all quantities
are given in units of $t$.

\subsection{Antiferromagnetism}

In the absence of any other symmetry breaking, the gap equation 
for the antiferromagnetic order parameter $A_{\bk}$ can be 
written as
\begin{equation}
 A_{\bk} = \frac{2}{L} \sum_{\bk'} U^{S}_{\bk\bk'} \,
 \frac{A_{\bk'}}{E^+_{\bk'} - E^-_{\bk'}} \,
 [f(E^+_{\bk'} - f(E^-_{\bk'})]
\end{equation}
with the two quasi-particle energy branches
\begin{equation}
 E^{\pm}_{\bk} = 
 \frac{\xi_{\bk} + \xi_{\bk+\bQ}}{2} \pm 
 \sqrt{\frac{1}{4}(\xi_{\bk} - \xi_{\bk+\bQ})^2 + A_{\bk}^2} \; .
\end{equation}
For a dispersion relation due to nearest and next-to-nearest 
neighbor hopping one has 
\begin{equation}
 E^{\pm}_{\bk} = 
 \eps^{t'}_{\bk} \pm 
 \sqrt{(\eps^t_{\bk})^2 + A_{\bk}^2} \, - \mu \; ,
\end{equation}
where $\eps^t_{\bk} = -2t(\cos k_x + \cos k_y)$ 
and $\eps^{t'}_{\bk} = -4t' \cos k_x \cos k_y \,$.

The antiferromagnetic order parameter affects the quasi-particle
energies most strongly near the magnetic Brillouin zone boundary, 
since $\eps^t_{\bk}$ vanishes there.
The dispersion of the two branches $E^{\pm}_{\bk}$ along this line 
is shown in Fig.~4 for a constant (momentum independent) $A_{\bk}$
and $t' < 0$.
\begin{figure}[ht]
\centering
\includegraphics[clip=true,width=6cm]{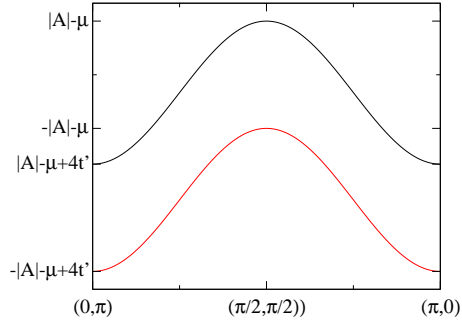}
\caption{(Color online)
 Dispersion of quasi-particle energies $E^{\pm}_{\bk}$  
 on the magnetic Brillouin zone boundary (''umklapp surface'') in 
 the antiferromagnetic state for a momentum independent $A_{\bk}$
 and $t'<0$.}
\label{fig4}
\end{figure}
Note that the momentum dependence of $A_{\bk}$ is indeed rather 
weak in our RG+MF calculations.
Within plain mean-field theory applied to the Hubbard model one has
$U^S_{\bk\bk'} = -U/2$ so that $A_{\bk}$ is completely momentum 
independent.
The antiferromagnetic mean-field theory has been applied to the 
two-dimensional Hubbard model in several earlier works.
\cite{LH,Sch90,CF,HV}

The two branches $E^+_{\bk}$ and $E^-_{\bk}$ are separated by a global 
energy gap if $|A_{\bk}| > 2|t'|$. If the chemical potential lies in that 
gap, such that $E^+_{\bk} > 0$ and $E^-_{\bk} < 0$ for all $\bk$, the
electron density is at half-filling and the system does not exhibit
any gapless excitations.
Gapless excitations do exist if $E^+_{\bk}$ or $E^-_{\bk}$ has zeros.
The equations $E^{\pm}_{\bk} = 0$ define effective Fermi surfaces of the
antiferromagnetic state. From Figs.\ 4 and 5 one can see that for $t'<0$ 
and close to half-filling the effective Fermi surfaces enclose hole pockets
in the branch $E^-_{\bk}$ around $(\pi/2,\pi/2)$ and electron pockets in
$E^+_{\bk}$ around $(\pi,0)$ and $(0,\pi)$.
\begin{figure}[ht]
\centering
\includegraphics[clip=true,width=7cm]{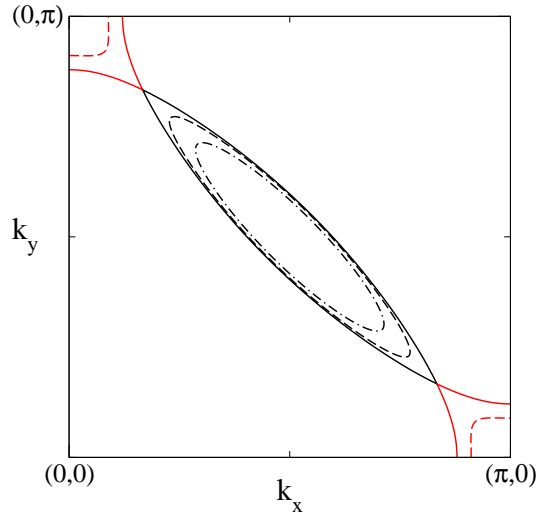}
\caption{(Color online)
 Effective Fermi surfaces in the antiferromagnetic state for 
 $t'=-0.2$, $\mu=-0.6$, and various momentum independent choices of 
 the antiferromagnetic gap function $A_{\bk} \,$: 
 0 (solid line), 0.15 (dashed line), 0.3 (dashed-dotted line).}
\label{fig5}
\end{figure}
Similar effective Fermi surfaces are obtained in a d-density wave
state.\cite{GZ}
For a momentum independent antiferromagnetic gap function $A_{\bk}$
the separatrices between states with hole pockets, electron pockets
and fully gapped states are given by simple linear combinations of 
the parameters $t'$, $\mu$ and $A$. For a fixed $t'<0$ the various
regimes in the plane spanned by $\mu$ and $A$ are shown in Fig.~6.
\begin{figure}[ht]
\centering
\includegraphics[clip=true,width=8cm]{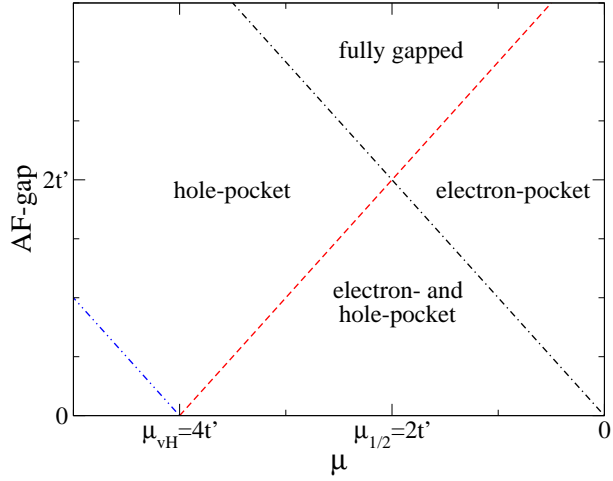}
\caption{(Color online)
 Topology of the effective Fermi surfaces in the plane
 spanned by the chemical potential $\mu$ and the antiferromagnetic
 gap $A$. No Fermi surface exists in the fully gapped regime.}
\label{fig6}
\end{figure}
Note the special point given by $\mu_{1/2} = A = 2t'$ at which the
separatrices for electron and hole pockets cross each other. 
For $\mu < \mu_{\rm vH}$ and a small $A_{\bk}$ the effective
Fermi surface is closed around $(0,0)$ and does not intersect with 
the magnetic Brillouin zone boundary.

A typical mean-field result for the $\mu$-dependence of the 
antiferromagnetic gap in the Hubbard model is shown in Fig.~7.
\begin{figure}[ht]
\centering
\includegraphics[clip=true,width=8cm]{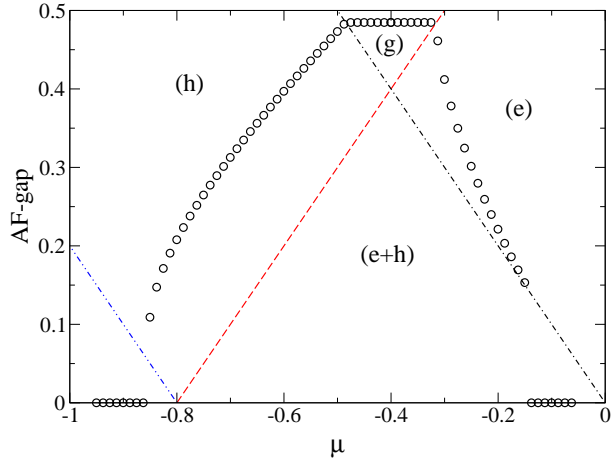}
\caption{(Color online)
 Mean-field solution for the antiferromagnetic gap as
 a function of $\mu$ for the Hubbard model with $t'=-0.2$ and
 $U=2.25$. Regions with different Fermi surface topology are
 indicated as in Fig.~\ref{fig6}.}
\label{fig7}
\end{figure}
The coupling strength $U$ has been chosen sufficiently large
to stabilize fully gapped half-filled solutions for $\mu$ near
$\mu_{1/2}$. The antiferromagnetic order parameter vanishes
discontinuously at the edges of the antiferromagnetic region. 
The effective Fermi surface of the antiferromagnetic states
consists (exclusively) of hole pockets for hole doping and of 
electron pockets for electron doping.
For other finite values of $t'$ the $\mu$-dependence of $A_{\bk}$
is qualitatively the same for sufficiently large $U$.
 
For $t'=0$ half-filled fully gapped antiferromagnetic solutions
are stabilized for any (arbitrarily small) $U$, but no stable
antiferromagnetic solutions away from half-filling exist. 
Solutions of the self-consistency equations leading to densities
away from half-filling correspond to maxima (instead of minima)
in the free energy and are therefore physically irrelevant.
\cite{Rei,GRH} 
If a certain density near half-filling is enforced, the system 
will thus phase separate in a half-filled antiferromagnetic and 
a non-half-filled paramagnetic region.

We now turn to results obtained from the combined RG+MF theory
described in the preceeding sections.
The effective interaction $U^S_{\bk\bk'}$ driving the
antiferromagnetic order is computed from the RG flow integrated 
down to a cutoff scale $\Lam_{\rm MF}$, and the momentum space 
entering the mean-field equations is correspondingly restricted 
by this cutoff.
Since the mode elimation is done only approximately both above
and below $\Lam_{\rm MF}$, the choice of $\Lam_{\rm MF}$ will
affect the accuracy of the final results.
In Fig.~8 we show the dependence of the antiferromagnetic gap
function $A_{\bk}$ on $\Lam_{\rm MF}$ for the perfect nesting
case ($t'=0$ at half-filling) with a relatively weak bare
coupling strength $U=2$. The different curves correspond to
momenta $\bk$ in six different patches interpolating between 
the $k_x$ or $k_y$ axis (patch 1) and the Brillouin zone diagonal 
(patch 6), see also Fig.~2. The patch-dependence is relatively
weak, reflecting the weak momentum dependence of $A_{\bk}$.
\begin{figure}[ht]
\centering
\includegraphics[clip=true,width=8cm]{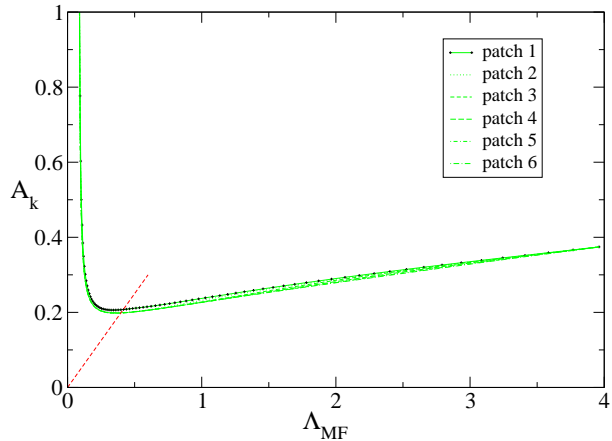}
\caption{(Color online)
 Antiferromagnetic gap as a function of the cutoff 
 $\Lam_{\rm MF}$ for the Hubbard model with pure nearest
 neighbor hopping ($t'=0$) and $U=2$ at half-filling. 
 Due to the weak momentum dependence of $A_{\bk}$ the 
 curves corresponding to different patches are very close 
 to each other.
 The intersection of the largest $A_{\bk}$ (patch 1) with
 the straight line given by $\Lam_{\rm MF} = 2 A_{\bk}$ yields
 at suitable choice of $\Lam_{\rm MF}$ leading to a 
 reasonable estimate of the true gap size.}
\label{fig8}
\end{figure}
Fluctuations captured by the one-loop RG flow reduce the
size of $A_{\bk}$ compared to the plain mean-field result
(corresponding to $\Lam_{\rm MF} = \Lam_0 = 4$).
This reduction is expected and can also be obtained by a
suitable perturbation theory for the symmetry-broken state.
\cite{GY,Don}
Within mean-field theory, the effective interaction is
driven exclusively by one-loop diagrams in the antiferromagnetic
particle-hole channel, leading to an enhancement compared to
the bare interaction. In the full one-loop flow other
contributions, especially from the particle-particle channel,
reduce this enhancement, leading to a smaller $A_{\bk}$.

The size of $A_{\bk}$ saturates at low $\Lam_{\rm MF}$
until it is pushed upwards by a strong divergence at a cutoff 
scale of the order of the gap size just before the upturn.
This divergence reflects the breakdown of the one-loop 
approximation at the energy scale of symmetry breaking.
The renormalized interaction becomes strong at that scale and
the influence of the antiferromagnetic order parameter on the 
flow becomes crucial. 
The latter is captured by the mean-field theory, which is 
definitely a better approximation at scales of the order of 
the final gap.
The one-loop flow without order parameter feedback fails even
in pure mean-field models at the scale where symmetry breaking
sets in.
Hence we have to stop the one-loop flow at a scale safely above
the scale of symmetry breaking. 
On the other hand one would like to capture the one-loop 
fluctuations not treated in mean-field theory down to the lowest 
possible scales.
Unfortunately there is no unique choice of an ''optimal'' 
$\Lam_{\rm MF}$.
In the following we will stop the one-loop flow at a scale
$\Lam_{\rm MF}$ which is twice as big as the largest (as a
function of $\bk$) gap value in the symmetry-broken state 
obtained from the mean-field theory for the states below 
$\Lam_{\rm MF}$. The relation $\Lam_{\rm MF} = 2 A_{\bk}$
corresponds to the straight line in Fig.~8. The estimate for
$A_{\bk}$ is obtained from the intersection of that line with
the largest $A_{\bk}$, that is, the ones with $\bk$ on patch 1.
In the present case this estimate is rather robust with respect
to slight shifts of $\Lam_{\rm MF}$.

\subsection{d-wave superconductivity}

It is well known that the one-loop RG flow generates an attractive
interaction in the d-wave Cooper channel.\cite{ZS,HM00a,HSFR,KK03a}
Although antiferromagnetic fluctuations contribute significantly to
this attraction, the Cooper instability ultimately competes with
antiferromagnetism.
In Fig.~9 we show results for the superconducting gap function
$\Delta_{\bk}$ as a function of the cutoff $\Lam_{\rm MF}$ in the
RG+MF scheme. Different curves correspond to different patches, as in
Fig.~8. The parameters have been chosen such that the antiferromagnetic
interaction is too weak to drive antiferromagnetic order.
\begin{figure}[ht]
\centering
\includegraphics[clip=true,width=8cm]{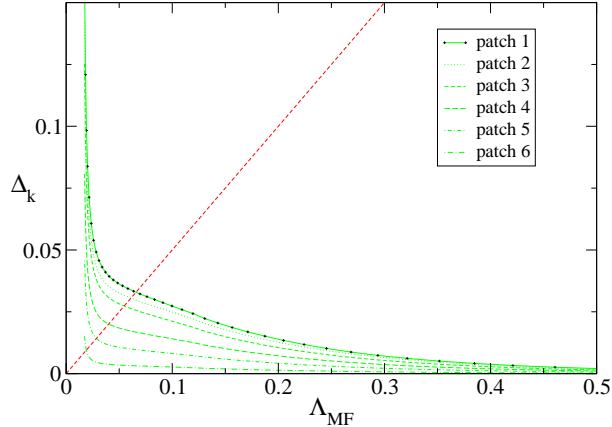}
\caption{(Color online) Superconducting gap as a function of 
 the cutoff $\Lam_{\rm MF}$ for the Hubbard model in a regime 
 where superconductivity is the only instability.
 Parameters: $U=2.5$, $t'=-0.2$, $\mu=-0.9265$.}
\label{fig9}
\end{figure}
Within plain mean-field theory no superconductivity is obtained since
the bare (repulsive) Hubbard interaction is repulsive in the s-wave
Cooper channel and vanishes for any other symmetry. 
Hence $\Delta_{\bk}$ vanishes for $\Lam_{\rm MF}$ at the band edge 
and increases monotonically upon lowering $\Lam_{\rm MF}$. The
pronounced momentum dependence of $\Delta_{\bk}$ reflects the
$d_{x^2-y^2}$ symmetry.
As in the case of antiferromagnetism, $\Delta_{\bk}$ diverges if 
the one-loop flow is continued below scales of the order of the
gap amplitude. A reasonable choice of $\Lam_{\rm MF}$ is obtained
from the condition $\Lam_{\rm MF} = 2\Delta_{\bk}$ (straight line
in Fig.~9) applied to the largest $\Delta_{\bk}$ (patch 1).
As long as one does not approach the divergence, the result for
$\Delta_{\bk}$ does not depend too much on the precise choice of 
$\Lam_{\rm MF}$.

\subsection{Antiferromagnetism and superconductivity}

We now analyze the competition between antiferromagnetism and
superconductivity within the full RG+MF theory, where both types
of order and also the possibility of their coexistence are allowed.
We focus on the case $t'<0$, which is realized in particular 
in the cuprate superconductors. 
We restrict ourselves to densities at and below half-filling.

In Fig.~10 we show results from the RG+MF calculation for the 
amplitudes of the d-wave superconducting gap $\Delta_{\bk}$ and
the antiferromagnetic gap $A_{\bk}$ as a function of the chemical 
potential $\mu$, and in Fig.~11 as a function of the electron
density. The latter is computed simultaneously with the order 
parameters and differs from the bare density corresponding to
$\mu$. 
\begin{figure}[ht]
\centering
\includegraphics[clip=true,width=8cm]{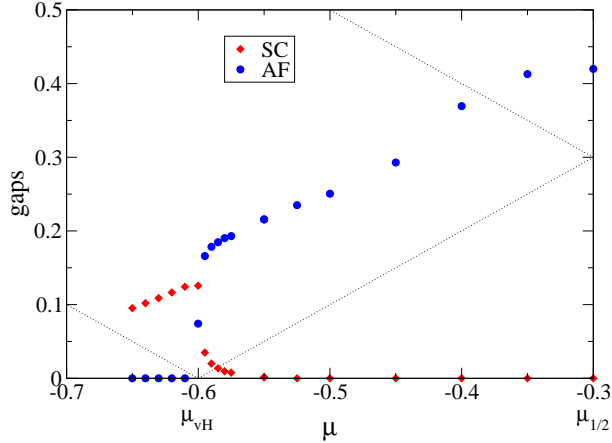}
\caption{(Color online)
 Amplitudes of the superconducting gap $\Delta_{\bk}$ and 
 the antiferromagnetic gap $A_{\bk}$ as a function of $\mu$ for 
 $U = 2.5$ and $t' = -0.15$. The dotted lines are the
 separatrices between different Fermi surface topologies as 
 specified in Fig.~6.}
\label{fig10}
\end{figure}
\begin{figure}[ht]
\centering
\includegraphics[clip=true,width=8cm]{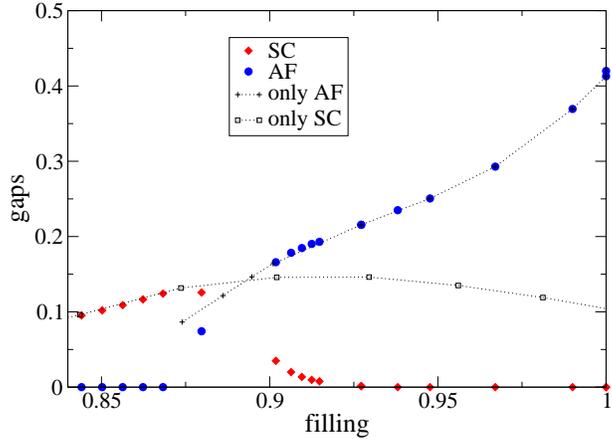}
\caption{(Color online)
 Amplitudes of $\Delta_{\bk}$ and $A_{\bk}$ as a function 
 of density for $U = 2.5$ and $t' = -0.15$.
 Filled colored symbols represent the results obtained for the
 combined theory with two order parameters, while in the results
 represented by open symbols only one order parameter, either
 antiferromagnetic or superconducting, was allowed in the mean-field
 calculation. 
 The two slightly different results for the antiferromagnetic gap at
 half-filling correspond to two different choices of $\mu$ within the 
 gap; this $\mu$-dependence is an artefact of the approximations.}
\label{fig11}
\end{figure}
The interaction $U=2.5$ is strong enough to stabilize an 
antiferromagnetic insulator at half-filling, in spite of the
magnetic frustration induced by the next-to-nearest neighbor 
hopping $t'=-0.15$. The system is fully gapped at half-filling
and the superconducting order parameter is strictly zero there.
With decreasing filling the antiferromagnetic gap decreases
monotonically, while $\Delta_{\bk}$ remains numerically zero. 
For electron densities below one, holes appear first in pockets 
around $(\pi/2,\pi/2)$, which define a surface of low-energy 
excitations of the non-half filled system. 
In principle, this residual Fermi surface is always unstable
against superconductivity, due to the attractive interaction 
in the d-wave Cooper channel.
However, very close to half-filling, where the pockets are small,
the superconducting gap is tiny, since the d-wave attraction is 
very small near the Brillouin zone diagonal. 
When the hole pockets are large enough, roughly for $n < 0.92$, 
a sizeable superconducting gap develops, coexisting with a finite 
antiferromagnetic order parameter. 
When $\mu$ approaches $\mu_{\rm vH}$, the antiferromagnetic gap
decreases rapidly. This leads also to a very fast decrease of the
electron density (see Fig.~11) as a function of decreasing $\mu$.
The antiferromagnetic gap vanishes for $\mu < \mu_{\rm vH}$, 
while $\Delta_{\bk}$ remains finite and decreases monotonically 
with decreasing filling.
Note that the antiferromagnetic transition is not generically 
situated as close to $\mu_{\rm vH}$ as in Fig.~10.

The numerical analysis is rather involved in the coexistence 
region and we have not clarified the nature of the magnetic transition
(first or steep second order). The presence of the superconducting
gap makes the magnetic transition smoother, but whether this
effect is big enough to yield a continuous transition is not
clear from the numerical data. In any case, fluctuations which
have been neglected in the mean-field approximation are expected
to play an important role in the transition region and may even 
affect the order of the transition.

The respective results for each order parameter 
when the other one is set to zero are also shown in Fig.~11. 
When $A_{\bk}$ is set to zero, the superconducting gap 
$\Delta_{\bk}$ persists even at half filling. 
When $\Delta_{\bk}$ is set to zero, the antiferromagnetic order
parameter is enhanced in the coexistence region. 
In both cases a finite value for one order parameter leads to a 
suppression of the other. 
While these results have been obtained for a weak on-site repulsion, 
they are in line with the behavior at stronger coupling as obtained
by the variational Monte Carlo technique,\cite{GL} by 
cluster and cellular dynamical mean-field theory,\cite{LK,CK}
and by variational cluster approximations\cite{SLMT,AAPH} 
for the two-dimensional Hubbard model.
The regime of substantial coexistence of antiferromagnetic order 
and superconductivity obtained in those calculations 
is typically larger than in our results. However, in the cluster
calculations the amount of coexistence seems to depend sensitively on
the cluster size.\cite{SLMT}
In a recent cellular dynamical mean-field calculation with a 2x2 cluster
it was found that coexistence is not possible if $U$ exceeds the band 
width.\cite{CK}

In Fig.~12 we show the angular dependence of $\Delta_{\bk}$ and
$A_{\bk}$ for three selected densities, along with the
corresponding effective Fermi surfaces.
The angle $\phi$ is defined with respect to the $k_x$ axis.
\begin{figure}[ht]
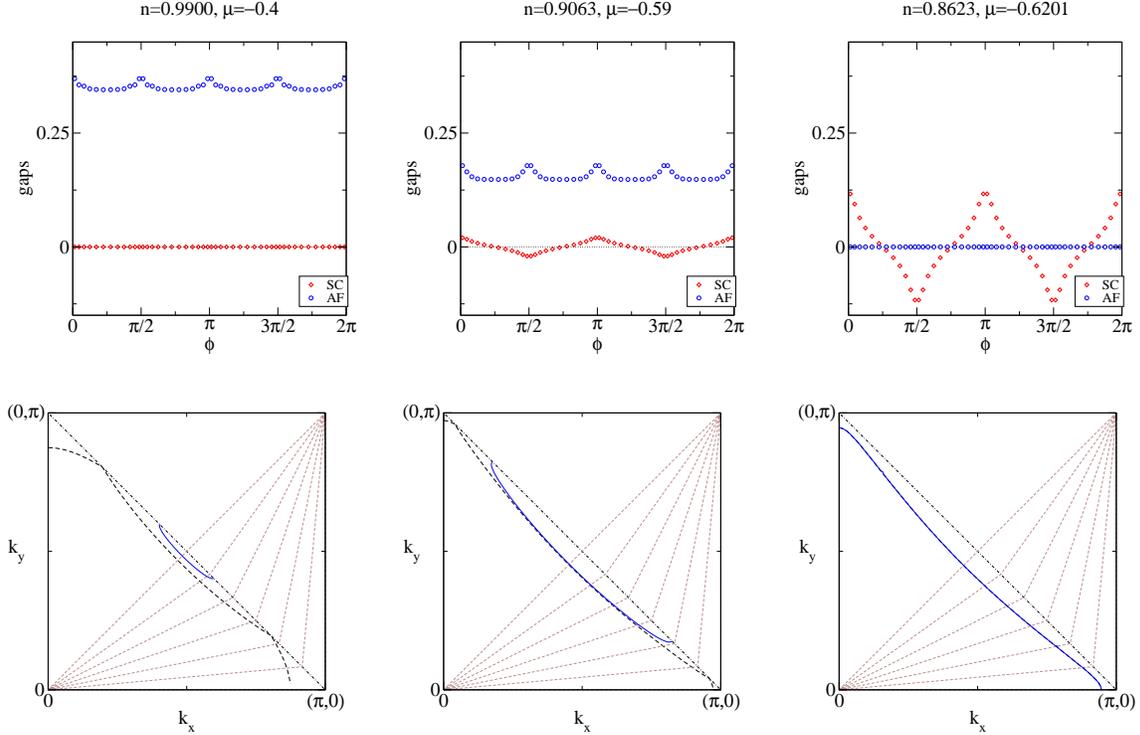

\includegraphics[width = 4.5cm]{fig12a.eps} \hspace{0.4cm}
\includegraphics[width = 4.5cm]{fig12b.eps} \hspace{0.4cm}
\includegraphics[width = 4.5cm]{fig12c.eps} \hfill 
\\
\phantom{.}
\includegraphics[width = 4.5cm]{fig12d.eps} \hspace{0.5cm}
\includegraphics[width = 4.5cm]{fig12e.eps} \hspace{0.5cm}
\includegraphics[width = 4.5cm]{fig12f.eps} \hfill
\caption{(Color online)
 {\em Top}: Angular dependence of the d-wave superconducting 
 and antiferromagnetic order parameters at three different values of 
 the chemical potential.
 {\em Bottom}: Fermi surfaces in the magnetic Brillouin zone 
 (solid blue lines); in the antiferromagnetic state close to 
 half-filling the Fermi surface forms a hole pocket around 
 $(\pi/2,\pi/2)$.
 The corresponding bare Fermi surfaces, backfolded with respect
 to Umklapp surface, are shown as broken lines.
 The straight lines indicate the patching scheme.
 The parameters $U=2.5$ and $t'=-0.15$ are the same as in 
 Figs.~10 and 11.}
\label{fig12}
\end{figure}
For $n=0.99$ we obtain a sizable antiferromagnetic gap which has
s-wave symmetry and is slightly anisotropic. The superconducting
gap is practically zero. 
The hole pocket enclosed by the effective Fermi surface is rather
small and does not support a sizable superconducting gap. 
For $n=0.906$ we observe a coexistence of both order parameters.
While $A_{\bk}$ is reduced in size compared to the case $n=0.99$,
its shape remains essentially the same. The hole pocket extends
further away from the Brillouin zone diagonal and allows for a
substantial superconducting gap with d-wave symmetry. In this
situation the low-energy excitations are gapped due to $A_{\bk}$
near the points $(\pi,0)$ and $(0,\pi)$, and due to $\Delta_{\bk}$
along the hole pocket, with nodes along the Brillouin zone diagonal.
For the even smaller density $n=0.862$ the antiferromagnetic order
parameter vanishes and only a d-wave superconducting gap remains,
which extends over the whole Fermi surface except at the nodal
points. Note that the momentum dependence of the gap function
has peaks near the van Hove points and is relatively flat near
the Brillouin zone diagonal, implying that terms beyond the lowest
d-wave harmonic $\cos k_x - \cos k_y$ contribute significantly
to $\Delta_{\bk}$.

In Figs.\ 10-12 the coupling strength $U$ has been chosen such 
that three different states, that is, antiferromagnetic insulator,
doped antiferromagnet coexisting with superconductivity, and a
pure d-wave superconductor could be obtained for different 
choices of $\mu$. 
By contrast, for a sufficiently small $U$ (at finite $t'$) 
the system is a d-wave superconductor for any $\mu$. 
On the other hand, for large $U$ the system is an antiferromagnet 
near half-filling and switches to a purely superconducting
state if the chemical potential descends below a certain critical
value. 
The superconducting gap in the doped antiferromagnet remains tiny
for large $U$.
A global view of the $U$ and $\mu$-dependences is presented
in Fig.~13, where we show the ground state phase diagram of the 
two-dimensional Hubbard model in the $\mu$-$U$ plane for a fixed 
$t'=-0.1$.
Note that the parameter region supporting substantial coexistence
of antiferromagnetism and superconductivity is rather narrow.
For $U=1.75$ and $U=2$, antiferromagnetic order is stabilized only
{\em below}\/ half-filling.
\begin{figure}[ht]
\centering
\includegraphics[clip=true,width=8cm]{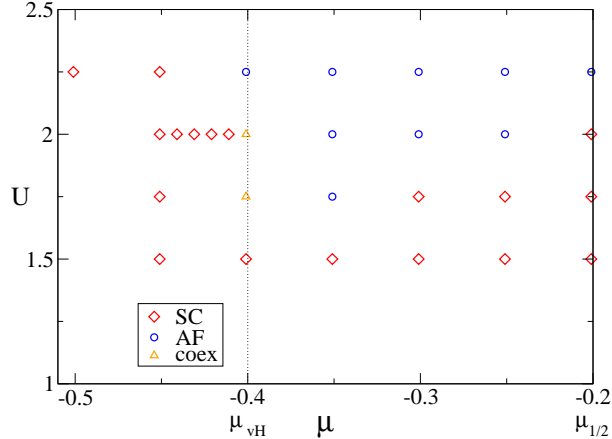}
\caption{(Color online)
 Points with antiferromagnetism, superconductivity and
 coexistence of both in the $\mu$-$U$ plane for a fixed $t'=-0.1$;
 ''coexistence'' means that the amplitude of $\Delta_{\bk}/A_{\bk}$
 is at least $10^{-3}$.}
\label{fig13}
\end{figure}

\section{Conclusion}

We have analyzed the competition between antiferromagnetism and
superconductivity in the two-dimensional Hubbard model by combining
a functional RG flow with a mean-field theory for symmetry-breaking. 
Effective interactions were computed by integrating out states
with an energy above a scale $\Lam_{\rm MF}$ via a one-loop RG
flow, which captures in particular the generation of an attractive
interaction in the d-wave Cooper channel from fluctuations in the
particle-hole channel.\cite{ZS,HM00a,HSFR}
These effective interactions were then used as an input for a 
mean-field treatment of the remaining states, below the scale
$\Lam_{\rm MF}$, with commensurate antiferromagnetism, singlet
superconductivity, and triplet $\pi$-pairing as the possible order
parameters.
Triplet $\pi$-pairing appears generically when antiferromagnetism
and singlet superconductivity coexist.\cite{PF} 
Our theory extends previous mean-field treatments of antiferromagnetism
and d-wave superconductivity, where the effective interactions were 
specified by a relatively simple ansatz with a few input parameters,
or they were identified with bare interactions of a microscopic model.
\cite{IDHR,MF,Kyu,YK2}
In the present work the size and shape (momentum dependence) of the 
effective interactions were actually \emph{computed}, within an 
approximation that captures fluctuations and is controlled at weak 
coupling.

It turned out that the feedback of $\pi$-pairing on the other
order parameters is negligble for the two-dimensional Hubbard model.
The key players, antiferromagnetism and d-wave superconductivity,
strongly compete. 
For a sufficiently large $U$ (depending on the size of $t'$) an
antiferromagnetic insulator is stabilized at half-filling, as
expected. 
Doping the antiferromagnet with holes (for $t'<0$) leads to an 
effective Fermi surface with hole pockets around $(\pi/2,\pi/2)$.
In principle, these give rise to a superconducting instability,
but the corresponding gap is usually tiny. 
At larger doping the antiferromagnetism breaks down and d-wave
superconductivity prevails. There is a small range of densities 
where both orders can coexist with a sizable order parameter 
for each.

Coexistence of antiferromagnetism and superconductivity in the 
two-dimensional Hubbard model has also been obtained by 
the variational Monte Carlo technique,\cite{GL} by 
cluster and cellular dynamical mean-field theory,\cite{LK,CK}
and by variational cluster approximations.\cite{SLMT,AAPH}
Coexistence has also been observed in some cuprate 
high-temperature superconductors in the underdoped regime.
\cite{SUX,MAX}

We have restricted our analysis of symmetry breaking to 
superconductivity and commensurate antiferromagnetism, while in
principle also other instabilities may arise. The most serious
candidates (and difficult to treat) are probably the various 
possibilities of incommensurate magnetic order in the low doping 
regime.\cite{CF,Sch90}
For the two-dimensional t-J model with a finite $t'$ it was shown 
that for very low doping a spiral phase is stabilized and that 
superconductivity develops in the spiral state.\cite{SK} 
Another possibility is the formation of a d-density wave, that 
is, a charge density wave with a wave vector near $(\pi,\pi)$ 
and a form factor with d-wave symmetry.\cite{CZ,CLMN} 
This ordered phase is known to occur in a suitable large $N$
extension of the t-J model, where it competes and partially 
coexists with d-wave superconductivity.\cite{CZ}
Renormalization group calculations for the two-dimensional
Hubbard model yield enhanced d-density wave correlations, but
the antiferromagnetic and superconducting instabilities are
stronger.\cite{HSR}

A more complete analysis of the two-dimensional Hubbard model by
renormalization group methods, including also order parameter
fluctuations at low energy scales, can be expected to yield further 
important clues for a better understanding of the interplay of 
magnetism and superconductivity.

\acknowledgments

We thank R.\ Gersch, C.\ Honerkamp, A.\ Katanin, M.\ Salmhofer, 
P.\ Strack, O.\ Sushkov, and R.\ Zeyher for valuable discussions.

\appendix

\section{Derivation of gap equations}

Let $\phi_{\bk}$ be any of the expectation values
$\bra n_{\bk\sg} \ket$, $\bra n^{\pi}_{\bk\sg} \ket$, 
$\bra p_{\bk} \ket$, $\bra p^{\pi}_{\bk} \ket$, 
or their complex conjugates.
Applying the necessary condition for a minimum,
$\partial\Omega/\partial\phi_{\bk} = 0$, to $\Omega$ in
Eq.\ (\ref{Omega}) yields
\begin{equation}
 \frac{\partial K^c}{\partial\phi_{\bk}} + 
 \sum_{\bk'} \sum_{\lam} 
 \frac{\partial E^{\lam}_{\bk'}}{\partial\phi_{\bk}} \,
 f(E^{\lam}_{\bk'}) = 0
\end{equation}
The derivatives of the energy eigenvalues $E^{\lam}_{\bk'}$ 
can be expressed in terms of minors $\det(M^{\lam,jj'}_{\bk'})$
of $\cM^{\lam}_{\bk'}$ and matrix elements $(\cM_{\bk'})_{jj'}$
of $\cM_{\bk'}$ as
\begin{equation}
 \frac{\partial E^{\lam}_{\bk'}}{\partial\phi_{\bk}} =
 \frac{1}{\sum_l \det \big( M^{\lam,ll}_{\bk'} \big)}
 \sum_{j,j'} 
 \frac{\partial (\cM_{\bk'})_{jj'}}{\partial\phi_{\bk}} \,
 (-1)^{j-j'} \, \det \big( M^{\lam,jj'}_{\bk'} \big) \; .
\end{equation}
This follows from acting with $\partial/\partial\phi_{\bk}$ 
on the equation for the eigenvalues
$\det \big( \cM_{\bk'} - E^{\lam}_{\bk'} {\bf 1} \big) = 0$
and the rule for derivatives of determinants
\begin{equation}
 \frac{\partial}{\partial\phi} \det(M) =
 \sum_{j,j'} \frac{\partial (M)_{jj'}}{\partial\phi} \,
 (-1)^{j-j'} \, \det(M^{jj'}) \; .
\end{equation}
The gap equations are now obtained by taking the specific
derivatives for each case.
Terms differing only by a momentum shift $\bQ$ can be 
summed up by using the identities 
$E^{\lam}_{\bk+\bQ} = E^{\lam}_{\bk}$ and
$\det \big( M^{\lam,jj'}_{\bk+\bQ} \big) =
 \det \big( M^{\lam,j+2,j'+2}_{\bk} \big)$, with
$j+2$ and $j'+2$ modulo 4, which follow from the block
structure (\ref{block}) of $\cM_{\bk}$. 

To derive Eq.~(\ref{denom}) we start from Kramer's rule
for matrix inversion in terms of minors,
$\det(M) \, (M^{-1})_{ll'} = (-1)^{l-l'} \det(M^{l'l})$.
Summing the diagonal elements yields
$\sum_l \det(M^{ll}) = \det(M) \, {\rm tr}(M^{-1})$.
Expressing the trace and the determinant by the
eigenvalues $m_{\alf}$ of $M$ yields
$\sum_l \det(M^{ll}) = 
 \sum_{\alf} \prod_{\alf' \neq \alf} m_{\alf'}$.
If one of the eigenvalues, say $m_{\lam}$, vanishes, only
the term $\alf = \lam$ contributes, yielding
$\sum_l \det(M^{ll}) = \prod_{\lam' \neq \lam} m_{\lam'}$.
Applying this identity to the matrix $\cM^{\lam}_{\bk}$ 
one obtains directly Eq.~(\ref{denom}).

\section{Attractive Hubbard model}

In the attractive Hubbard model away from half-filling only
the Cooper channel leads to an instability, namely s-wave
superconductivity.\cite{MRS}
All other channels yield just finite renormalizations. 
In Fig.~14 we show the dependence of the superconducting gap 
on $\Lam_{\rm MF}$ as obtained from the RG+MF theory for the 
attractive Hubbard model with $U=-1.5$ and $t'=-0.1$ at 
quarter-filling. 
\begin{figure}[ht]
\centering
\includegraphics[clip=true,width=8cm]{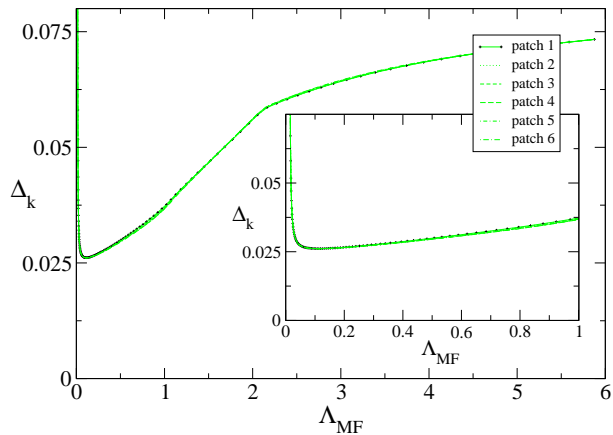}
\caption{(Color online)
 Superconducting gap $\Delta_{\bk}$ as a function of 
 the cutoff $\Lam_{\rm MF}$ for the attractive Hubbard model 
 ($U=-1.5$, $t'=-0.1$) at quarter-filling. Due to the
 weak momentum dependence of $\Delta_{\bk}$ curves 
 corresponding to different patches lie almost on top of
 each other. The inset shows $\Delta_{\bk}$ for
 $\Lam_{\rm MF} \leq 1$ with a higher resolution.}
\label{fig14}
\end{figure}
As in the case of antiferromagnetism in the repulsive model, 
the gap size is reduced by fluctuations compared to the mean-field 
result.
For $\Lam$ between $4 - \mu \approx 6$ and $4 + \mu \approx 2$ 
only states above the Fermi level are integrated out, 
while states below the Fermi level are captured only for 
smaller $\Lam$. This is the reason for the kink in $\Delta_{\bk}$ 
at $\Lam_{\rm MF} = 4 + \mu$. 
The gap diverges at a critical scale $\Lam_c$ due to the breakdown
of the one-loop approximation discussed already in Sec.~IV A.
For $\Lam_c \ll \Lam_{\rm MF} \ll 1$ the dependence of 
$\Delta_{\bk}$ on $\Lam_{\rm MF}$ is relatively weak.
This is the regime where the cutoff is still high enough so
that the one-loop approximation is still accurate, but on 
the other hand already sufficiently low so that the flow of 
the effective interaction $V_{\bk\bk'}$ is dominated by the 
particle-particle channel, such that interactions beyond the 
reduced (mean-field) model are irrelevant.


\vfill\eject


\begin{thebibliography}{99}
\bibitem{fn1} Spontaneous symmetry breaking is not always
 signalled by divergent interactions; in case of first 
 order transitions they usually remain finite.
\end{thebibliography}
\end{document}